\documentclass[12pt]{article}
\pdfoutput=1
\usepackage{amsmath}
\def\be{\begin{equation}}
\def\ee{\end{equation}}
\def\arr{\begin{array}{rll}}
\def\ea{\end{array}}
\def\bea{\begin{eqnarray}}
\def\eea{\end{eqnarray}}

\def\N2{$N{=}2$}

\def\>{\rangle}
\def\<{\langle}
\def\+{\dagger}
\def\={\ =\ }

\begin{document}
\renewcommand{\thefootnote}{\fnsymbol{footnote}}
\begin{titlepage}
\setcounter{page}{0}

\begin{center}
{\LARGE\bf $\mathcal{N}=3$ super--Schwarzian  }\\
\vskip 0.5cm
{\LARGE\bf  from $OSp(3|2)$ invariants}\\

\vskip 1.5cm
\textrm{\Large Anton Galajinsky
}
\vskip 0.7cm
{\it
Tomsk Polytechnic University,
634050 Tomsk, Lenin Ave. 30, Russia}

\vskip 0.2cm
{e-mail:
galajin@tpu.ru}
\vskip 0.5cm

\end{center}
\vskip 1cm
\begin{abstract} \noindent
It was recently demonstrated that the $\mathcal{N}=0,1,2,4$ super--Schwarzian derivatives can be constructed by applying the method of nonlinear realizations to the finite--dimensional (super)conformal groups
$SL(2,R)$, $OSp(1|2)$, $SU(1,1|1)$, and $SU(1,1|2)$, respectively. In this work, a similar scheme is realised for $OSp(3|2)$. It is shown that the $\mathcal{N}=3$ case exhibits a surprisingly richer structure of invariants, the $\mathcal{N}=3$ super--Schwarzian being a particular member. We suggest that the extra invariants may  prove useful in building an $\mathcal{N}=3$ supersymmetric extension of the Sachdev--Ye--Kitaev model.
\end{abstract}


\end{titlepage}

\renewcommand{\thefootnote}{\arabic{footnote}}
\setcounter{footnote}0

\noindent
{\bf 1. Introduction}\\

\noindent
A remarkable property of the super--Schwarzian derivatives is that they hold invariant under finite--dimensional superconformal transformations \cite{F,Cohn,Sch,MU}. Similarly to other conformal invariants, which are widely used within the context of the AdS/CFT--correspondence, they proved useful for holographic description of systems which exhibit many--body quantum chaos (see e.g. \cite{FGMS,MTV,BBN} and references therein).
Such objects are characterised by a specific composition law under the change of the argument and the fact that setting them to zero one reveals a finite--dimensional $\mathcal{N}$--extended superconformal transformation acting in the odd sector of $\mathcal{S}^{1|\mathcal{N}}$ superspace\footnote{Because conformal transformations in $\mathcal{R}^1$ involve the inversion $t \to \frac{1}{t}$,
the conformal group $SL(2,R)$ does not act globally on $\mathcal{R}^1$, but rather on $\mathcal{S}^1=\mathcal{RP}^1$.} \cite{F,Cohn,Sch,MU}.

The super--Schwarzian derivatives were originally introduced by physicists within the context of superconformal field theory. Computing a finite superconformal transformation of the super stress--energy tensor underlying a $2d$ $\mathcal{N}$--extended conformal field theory, one reveals the $\mathcal{N}$--extended super--Schwarzian as the anomalous term. Because for $\mathcal{N}\geq 5$ the construction of the central term operator is problematic \cite{ChK}, in modern literature it is customary to focus on the $\mathcal{N}=0,1,2,3,4$ cases only.

In a series of recent works \cite{AG1,AG,GK}, an alternative approach to building the $\mathcal{N}=0,1,2,4$ super--Schwarzian derivatives was proposed, which is based upon the method of nonlinear realizations \cite{CWZ} applied to the finite--dimensional (super)conformal groups $SL(2,R)$, $OSp(1|2)$, $SU(1,1|1)$, and $SU(1,1|2)$. The advantage of the scheme is that nether infinite--dimensional extensions of the supergroups, nor conformal field theory techniques, nor the analysis of central charges/cocycles are needed.

A finite--dimensional $1d$ superconformal algebra typically involves\footnote{For simplicity of the presentation, here we omit all indices carried by the generators.} even generators of translations (P), dilatations (D), special conformal transformations (K) and the $R$--symmetry transformations ($\mathcal{J}$), while odd generators include (complex) supersymmetry charges (Q) and their superconformal partners (S). The method in \cite{AG1,AG,GK} consists of four steps. First, each generator in the superalgebra is accompanied by a Goldstone superfield of the same Grassmann parity
\begin{eqnarray*}
\begin{array}{|l|r|r|r|r|c|}
\hline
P & D & K & \mathcal{J} & Q & S \\
\hline
\rho(t,\theta) & \nu(t,\theta) & \mu(t,\theta) & \lambda(t,\theta) & \psi(t,\theta) & \phi(t,\theta) \\
\hline
\end{array}
\end{eqnarray*}
and the conventional group--theoretic element $g=e^{{\rm i} \rho P} e^{\psi Q} e^{\phi S} e^{{\rm i} \mu  K} e^{{\rm i} \nu D} e^{{\rm i} \lambda \mathcal{J}}$ is introduced \cite{CWZ}. Then the odd analogues of the Maurer--Cartan one--forms are computed ${g}^{-1} \mathcal{D} g={\rm i} \omega_P P+{\rm i} \omega_D D+{\rm i} \omega_K K+{\rm i} \omega_{\mathcal{J}} \mathcal{J}+\omega_Q Q+\omega_S S$, where $\mathcal{D}$ is the covariant derivative, which give rise to the superconformal invariants $(\omega_P,\omega_D,\omega_K,\omega_{\mathcal{J}},\omega_Q,\omega_S)$ built in terms of the superfields $(\rho,\nu,\mu,\lambda,\psi,\phi)$. As the third step, one imposes the minimum set of constraints
\be
\omega_D=0, \qquad \omega_Q=\mbox{const}, \qquad \omega_S=0,
\nonumber
\ee
which allow one to express all the Goldstone superfields entering $g$ in terms of the fermionic superfield $\psi$. Finally, substituting $(\rho,\nu,\mu,\lambda,\phi)$ back into the superconformal invariants,
one either gets zero or reproduces the $\mathcal{N}=0,1,2,4$ super--Schwarzian derivative acting upon $\psi$ \cite{AG1,AG,GK}.

The goal of this note is to demonstrate that the $\mathcal{N}=3$ case exhibits a surprisingly richer structure, the $\mathcal{N}=3$ super--Schwarzian being a particular member of a larger set of $OSp(3|2)$ invariants involving $\psi$ alone. Such extra invariants may prove useful in building an $\mathcal{N}=3$ supersymmetric extension of the Sachdev--Ye--Kitaev model.

The work is organised as follows. In the next section, superconformal diffeomorphisms of $\mathcal{S}^{1|3}$ superspace are considered and conditions which follow from the requirement that the covariant derivative transforms homogeneously are analysed. The infinitesimal form of $OSp(3|2)$ transformations is presented as well. In Sect. 3, $OSp(3|2)$ invariants involving $\psi_i$ alone are constructed by applying the scheme outlined above. The $\mathcal{N}=3$ super--Schwarzian proves to be a particular member. Its properties are discussed in Sect. 4. We summarise our results and discuss possible further developments in the concluding Sect. 5.

Throughout the text, summation over repeated indices is understood.

\vspace{0.5cm}

\noindent
{\bf 2. Superconformal diffeomorphisms of $\mathcal{S}^{1|3}$}\\

\noindent
$\mathcal{S}^{1|3}$ superspace is parametrized by a real bosonic coordinate $t$ and its real
ferminic partner $\theta_i$, which carries a vector index $i=1,2,3$. It can be conveniently represented as the supergroup manifold
\be
g=e^{{\rm i} t h} e^{\theta_i q_i},
\ee
where even $h$ and odd $q_i$ obey the
$d=1$, $\mathcal{N}=3$ supersymmetry algebra
\be
\{q_i, q_j\}=2 h \delta_{ij}.
\ee
The left action of the supergroup on the superspace, $g'=e^{{\rm i} a h} e^{\epsilon_i q_i} \cdot g$, where $a$ and $\epsilon_i$ are even and odd supernumbers, respectively, generates the $d=1$, $\mathcal{N}=3$ supersymmetry transformations
\be
t'=t+a; \qquad  {\theta}'_i=\theta_i+\epsilon_i, \qquad t'=t+{\rm i} \epsilon_i \theta_i.
\ee
Covariant derivative, which anticommutes with the supersymmetry generator, reads
\be\label{covd}
{\mathcal{D}}_i=\partial_i-{\rm i} \theta_i \partial_t, \qquad \{{\mathcal{D}}_i, {\mathcal{D}}_j \}=-2 {\rm i} \delta_{ij} \partial_t,
\ee
where $\partial_t=\frac{\partial}{\partial t}$, $\partial_i=\frac{\vec{\partial}}{\partial \theta_i}$.

Superconformal diffeomorphisms of $\mathcal{S}^{1|3}$ are introduced as the transformations
\be\label{sdiff}
t'=\rho(t,\theta), \qquad \theta'_i=\psi_i (t,\theta),
\ee
where $\rho$ is a real bosonic superfield and $\psi_i$ is a real fermionic superfield,
under which the covariant derivative transforms homogeneously \cite{Sch}
\be\label{hom1}
{\mathcal{D}}_i=\left({\mathcal{D}}_i \psi_j\right) {\mathcal{D}}'_j.
\ee
Eq. (\ref{hom1}) implies the constraint
\be\label{Const1}
{\mathcal{D}}_i \rho+{\rm i} \psi_j {\mathcal{D}}_i \psi_j=0,
\ee
from which one gets further restrictions
\bea\label{Const2}
&&
\left( {\mathcal{D}}_i \psi_k \right) \left({\mathcal{D}}_j \psi_k\right)=\frac 13 \delta_{ij} {\mathcal{D}}\psi {\mathcal{D}}\psi , \qquad ~ \left({\mathcal{D}}_k \psi_i \right) \left({\mathcal{D}}_k \psi_j \right)=\frac 13 \delta_{ij} {\mathcal{D}}\psi {\mathcal{D}}\psi,
\nonumber\\[2pt]
&&
{\mathcal{D}}_i \left({\mathcal{D}}\psi {\mathcal{D}}\psi \right)=-6 {\rm i} {\dot\psi}_j {\mathcal{D}}_i  \psi_j, \qquad \qquad \dot\rho={\rm i} \psi_i {\dot\psi}_i+\frac 13 {\mathcal{D}}\psi {\mathcal{D}}\psi ,
\eea
where the dot designates the derivative with respect to $t$ and ${\mathcal{D}}\psi {\mathcal{D}}\psi :=\left({\mathcal{D}}_i \psi_j \right) \left( {\mathcal{D}}_i \psi_j \right)$. In particular, $\rho$ is fixed provided $\psi_i$ is known.

In what follows, we will need the explicit form of $\psi_i$ obeying (\ref{Const2}). Considering the component decomposition in the odd variables parameterizing the superspace
\be\label{sf}
\psi_i (t,\theta)=\alpha_i (t)+ \theta_a b_{a i} (t)+\frac 12 \theta_a \theta_b \beta_{a b i} (t)+\frac{1}{3!} \epsilon_{abc} \theta_a \theta_b \theta_c g_i (t),
\ee
where ($b_{a i}$, $g_i$) and ($\alpha_i$, $\beta_{a b i}$) are bosonic and fermionic functions of $t$, respectively, and $\epsilon_{abc}$ is the Levi-Civita symbol, and making use of the covariant projection method, in which components of a superfield are linked to its covariant derivatives evaluated at $\theta_i=0$, one gets
\bea\label{sf1}
b_{ij}=u(t) {\mbox{exp} (\tilde\xi)}_{ij}, \qquad \beta_{ijk}=\frac{3 {\rm i} }{b^2} \left( (b_{is} {\dot\alpha}_s) b_{jk}-(b_{js} {\dot\alpha}_s) b_{ik}\right), \qquad
g_i=\frac{1}{2 u(t)} \epsilon_{ijk} {\dot\alpha}_j {\dot\alpha}_k,
\eea
where $u(t)$ is an arbitrary bosonic function of $t$, ${\tilde\xi}_{ij}=\xi_k \epsilon_{kij}$ involve a real bosonic vector parameter $\xi_k$ such that ${\mbox{exp} (\tilde\xi)}_{ij}={\mbox{exp} (-\tilde\xi)}_{ji}$, and $b^2=b_{ij} b_{ij}$.
Note that similarly to ${\mathcal{D}}_i \psi_j$ the bosonic component $b_{ij}$ obeys the equations
\be
b_{ik} b_{jk}=\frac 13 \delta_{ij} b^2, \qquad b_{ki} b_{kj}=\frac 13 \delta_{ij} b^2,
\ee
which mean that the parameter $\xi_k$ above represents finite $SO(3)$ transformations.

Thus, solving the quadratic constraints (\ref{Const2}), one links all the components of $\psi_i$ to a bosonic function $u(t)$ and a fermionic vector function $\alpha_i (t)$.
The Taylor series expansions of $u(t)$ and $\alpha_i (t)$ involve an infinite number of constant parameters, which all together represent an infinite--dimensional extension of $OSp(3|2)$ supergroup.

In what follows, we will need the infinitesimal form of $OSp(3|2)$ transformations acting upon $\rho$ and $\psi$. They can be obtained by analogy with the $\mathcal{N}=0,1,2,4$ cases studied in \cite{AG1,AG,GK} and the final result reads
\begin{align}
&
\rho'=\rho+a, && \psi'_i=\psi_i;
\nonumber\\[6pt]
&
\rho'=\rho+b\rho, && \psi'_i=\psi_i+\frac 12 b \psi_i;
\nonumber\\[6pt]
&
\rho'=\rho+c \rho^2, && \psi'_i=\psi_i+c \rho \psi_i;
\nonumber
\end{align}
\begin{align}\label{tr}
&
\rho'=\rho, && \psi'_i=\psi_i-\epsilon_{ijk} \xi_j \psi_k;
\nonumber\\[6pt]
&
\rho'=\rho+{\rm i} \epsilon_i \psi_i, && \psi'_i=\psi_i+\epsilon_i;
\nonumber\\[6pt]
&
\rho'=\rho-{\rm i} \rho (\kappa_i \psi_i), && \psi'_i=\psi_i-\rho \kappa_i- {\rm i} (\kappa_j \psi_j) \psi_i.
\end{align}
Here the bosonic parameters $(a,b,c,\xi_i)$ correspond to translations, dilatations, special conformal transformations, and $SO(3)$ rotations, respectively, while the fermionic parameters $(\epsilon_i,\kappa_i)$ are associated with supersymmetry transformations and superconformal boosts.
Note that both the original and transformed superfields depend on the same arguments $(t,\theta)$ such that the transformations affect the form of the superfields only, e.g. $\delta \rho=\rho'(t,\theta)-\rho(t,\theta)$.
Computing the algebra of the infinitesimal transformations (\ref{tr}), one can verify that it does reproduce the structure relations of $osp(3|2)$ superalgebra (see Eq. (\ref{algebra}) below).

Given the infinitesimal transformations (\ref{tr}), one can readily obtain the relations
\bea
&&
{\mathcal{D}}\psi' {\mathcal{D}}\psi'=(1+b) {\mathcal{D}}\psi {\mathcal{D}}\psi, \qquad {\mathcal{D}}\psi' {\mathcal{D}}\psi'=(1+2c\rho) {\mathcal{D}}\psi {\mathcal{D}}\psi,
\nonumber\\[2pt]
&&
{\mathcal{D}}\psi' {\mathcal{D}}\psi'=(1-2{\rm i} (\kappa_i \psi_i)) {\mathcal{D}}\psi {\mathcal{D}}\psi,
\eea
which will prove useful in what follows.

\vspace{0.5cm}

\noindent
{\bf 3. $OSp(3|2)$ invariants involving $\psi_i$} alone\\

\noindent
Let us apply the method of nonlinear realizations \cite{CWZ} to $OSp(3|2)$ with the aid to obtain superconformal invariants involving the fermionic superfield $\psi_i$ alone. Later on, the latter will be identified with the argument of the $\mathcal{N}=3$ super--Schwarzian.

As the first step, one considers
the structure relations of $osp(3|2)$ superalgebra
\begin{align}\label{algebra}
&
[P,D]={\rm i} P, && [P,K]=2{\rm i} D,
\nonumber\\[10pt]
&
[D,K]={\rm i} K, && [\mathcal{J}_i,\mathcal{J}_j]={\rm i} \epsilon_{ijk} \mathcal{J}_k,
\nonumber\\[10pt]
&
\{ Q_i, Q_j \}=2P \delta_{ij}, &&
\{ Q_i, S_j \}=\epsilon_{ijk} \mathcal{J}_k-2\delta_{ij} D,
\nonumber\\[2pt]
&
\{ S_i, S_j \}=2K \delta_{ij}, && [D,Q_i] = -\frac{{\rm i}}{2} Q_i,
\nonumber\\[2pt]
& [D,S_i] =\frac{{\rm i}}{2} S_i, && [K,Q_i] ={\rm i} S_i,
\nonumber\\[2pt]
&
[P,S_i]=-{\rm i} Q_i, && [\mathcal{J}_i,Q_j] ={\rm i} \epsilon_{ijk} Q_k,
\nonumber\\[10pt]
&
[\mathcal{J}_i,S_j] ={\rm i} \epsilon_{ijk} S_k. &&
\end{align}
Here $(P,D,K,\mathcal{J}_i)$ are (Hermitian) bosonic generators of translations, dilatations, special conformal transformations, and $SO(3)$ rotations, respectively, while $Q_i$ and $S_i$ are (Hermitian) fermionic generators of supersymmetry transformations and superconformal boosts.

Then each generator in the superalgebra is accompanied by a Goldstone superfield of the same Grassmann parity and the group--theoretic element is introduced
\be\label{elem}
g=e^{{\rm i} t h} e^{\theta_i q_i} e^{{\rm i} \rho P} e^{\psi_i Q_i} e^{\phi_i S_i} e^{{\rm i} \mu  K} e^{{\rm i} \nu D} e^{{\rm i} \lambda_i \mathcal{J}_i},
\ee
where $(\rho,\mu,\nu,\lambda_i)$ are real bosonic superfields and $(\psi_i,\phi_i)$ are real fermionic superfields. In what follows, $\rho$ and $\psi_i$ are identified with those in the preceding section and the constraints (\ref{Const1}), (\ref{Const2}) are assumed to hold. Note that such a choice of $g$ is suggested by the previous studies of $d=1$, $\mathcal{N}$--extended conformal mechanics in \cite{IKL,IKL1}.

Afterwards, one uses ${g}$ and the covariant derivative (\ref{covd}) so as to to build the odd analogues of the Maurer--Cartan one--forms
\be
{g}^{-1} \mathcal{D}_i g={\rm i} {\left(\omega_D\right)}_i D+{\rm i} {\left(\omega_K\right)}_i K+{\left( \omega_Q \right)}_{ij}  Q_j+{\left( \omega_S \right)}_{ij}  S_j+{\rm i} {\left( \omega_{\mathcal{J}} \right)}_{ij} \mathcal{J}_j-q_i,
\ee
where
\bea\label{invariants}
&&
{\left(\omega_D\right)}_i=\mathcal{D}_i \nu-2i \left(\mathcal{D}_i \psi_j\right) \phi_j,
\nonumber\\[8pt]
&&
{\left(\omega_K\right)}_i=e^\nu \left(\mathcal{D}_i \mu+2 {\rm i} \mu \left(\mathcal{D}_i \psi_j\right) \phi_j+{\rm i} \left(\mathcal{D}_i \phi_j\right) \phi_j \right),
\nonumber\\[8pt]
&&
{\left( \omega_Q \right)}_{ij}=e^{-\frac{\nu}{2}} \left(\mathcal{D}_i \psi_k \right) {\mbox{exp} (\tilde\lambda)}_{kj}
\nonumber\\[2pt]
&&
{\left( \omega_S \right)}_{ij}=e^{\frac{\nu}{2}} \left(\mathcal{D}_i \phi_k+\mu \mathcal{D}_i \psi_k +{\rm i} \left(\mathcal{D}_i \psi_l \right) \phi_l \phi_k \right) {\mbox{exp} (\tilde\lambda)}_{kj},
\nonumber\\[2pt]
&&
{\rm i} {\left( \omega_{\mathcal{J}} \right)}_{ij} \mathcal{J}_j=
e^{-{\rm i} \lambda_k \mathcal{J}_k} \left(\mathcal{D}_i e^{{\rm i} \lambda_p \mathcal{J}_p} \right)
-\epsilon_{jkl} \left(\mathcal{D}_i \psi_j\right) \phi_k
\left( e^{-{\rm i} \lambda_s \mathcal{J}_s} \mathcal{J}_l e^{{\rm i} \lambda_r \mathcal{J}_r}\right),
\eea
with ${\tilde\lambda}_{ij}=\lambda_k \epsilon_{kij}$. The invariant ${\left(\omega_P\right)}_i$ turns out to vanish identically as a consequence of the constraint (\ref{Const1}) imposed above.

By construction, ${g}^{-1} \mathcal{D}_i g$ hold invariant under the transformation $g \to \tilde g\cdot g$ with $\tilde g \in OSp(3|2)$ and, hence, Eqs. (\ref{invariants}) describe invariants of $OSp(3|2)$. These can be used to eliminate all the superfields except $\rho$ and $\psi_i$ from the consideration as well as to provide invariants involving $\psi_i$ alone. Within the method of nonlinear realizations, imposing constraints is attributed to the inverse Higgs phenomenon \cite{IO}.

Guided by our previous studies in \cite{AG,GK}, let us impose the following constraints\footnote{Had we not chosen (\ref{Const1}) earlier, it could have been imposed here by setting ${\left(\omega_P\right)}_i=0$.
}
\be\label{const3}
{\left(\omega_D\right)}_i=0, \qquad {\left( \omega_S \right)}_{ij}=0, \qquad {\left( \omega_Q \right)}_{ij}=r_{ij},
\ee
where $r_{ij}$ is a real matrix with even supernumber elements (coupling constants).\footnote{The consistency requires $r_{ij}$ to obey the condition $r_{ik} r_{jk}=\frac 13 \delta_{ij} r^2$ with $r^2= (r_{ij} r_{ij})$ (see Eq. (\ref{La})).} Using the last condition and the identity ${\mbox{exp} (\tilde\lambda)}_{ij}={\mbox{exp} (-\tilde\lambda)}_{ji}$, one can express $\nu$ in terms of $\psi_i$
\be\label{nu}
e^\nu=\frac{{\mathcal{D}}\psi {\mathcal{D}}\psi }{r^2}.
\ee
Computing the covariant derivative of (\ref{nu}) and taking into account ${\left(\omega_D\right)}_i=0$ and (\ref{Const2}), one links $\phi_i$ to $\psi_i$
\be
\phi_i=-\frac{3 \partial_t \psi_i}{ {\mathcal{D}}\psi {\mathcal{D}}\psi }.
\ee
Contracting ${\left( \omega_S \right)}_{ij}$ with ${\left( \omega_Q \right)}_{ij}$, one then relates $\mu$ to $\psi_i$
\be
\mu=-\frac 32 \partial_t \left(\frac{1}{{\mathcal{D}}\psi {\mathcal{D}}\psi }\right),
\ee
while ${\left( \omega_Q \right)}_{ij}=r_{ij}$ determines $\lambda_i$
\be\label{La}
{\mbox{exp} (\tilde\lambda)}_{ij}=\frac{3 e^{\frac{\nu}{2}} \left({\mathcal{D}}_k \psi_i \right) r_{kj}}{{\mathcal{D}}\psi {\mathcal{D}}\psi}.
\ee

Substituting $(\mu,\nu,\lambda_i,\phi_i)$ back into Eqs. (\ref{invariants}) and choosing for definiteness a regular representation for $\mathcal{J}_l$ entering $\left( \omega_{\mathcal{J}} \right)$, i.e. ${\left(\mathcal{J}_i\right)}_{jk}=- {\rm i} \epsilon_{ijk}$, one finally gets $OSp(3|2)$ invariants that involve $\psi_i$ alone
\bea\label{INV}
&&
\frac{1}{\mathcal{D} \psi \mathcal{D} \psi }\left(\frac 12 (\mathcal{D}_k \psi_l) [\mathcal{D}_i, \mathcal{D}_j] \psi_l-{\rm i} \delta_{ki} \left({\dot\psi}_l \mathcal{D}_j \psi_l\right)+{\rm i} \delta_{kj} \left({\dot\psi}_l \mathcal{D}_i \psi_l \right)  \right), \qquad ({\mathcal{D}}_i \psi_j) { \left(\frac{{\dot\psi}_j}{{\mathcal{D}}\psi {\mathcal{D}}\psi} \right)}^{\cdot},
\nonumber\\[2pt]
&&
\frac{-6 {\rm i}}{\mathcal{D} \psi \mathcal{D} \psi }\left( (\mathcal{D}_i {\dot\psi}_l) ( \mathcal{D}_j \psi_l)-(\mathcal{D}_j {\dot\psi}_l) (\mathcal{D}_i \psi_l) +\frac{6 {\rm i}( {\dot\psi}_k \mathcal{D}_i \psi_k ) ( {\dot\psi}_l \mathcal{D}_j \psi_l )}{\mathcal{D} \psi \mathcal{D} \psi}\right)=
\nonumber\\[2pt]
&&
\qquad
=[\mathcal{D}_i,\mathcal{D}_j] \ln{(\mathcal{D} \psi \mathcal{D} \psi)}+(\mathcal{D}_i \ln{(\mathcal{D} \psi \mathcal{D} \psi)}) (\mathcal{D}_j \ln{(\mathcal{D} \psi \mathcal{D} \psi)})-\frac{6 {\rm i} {\dot\psi}_l [\mathcal{D}_i,\mathcal{D}_j] \psi_l }{\mathcal{D} \psi \mathcal{D} \psi}.
\eea
To the best of our knowledge, these expressions are new and have not yet been presented in the literature.

One may wonder whether the invariants in (\ref{INV}) are functionally dependent. The clue is the dimension analysis. As $\psi_i$ and $\mathcal{D}_i$  have opposite dimensions ($(\mathcal{D} \psi \mathcal{D} \psi)$ is dimensionless), one may expect to find a relation between the third invariant and the covariant derivative of the first invariant. Similarly, the second invariant might, in principle, be related to the covariant derivative of the third invariant and the double covariant derivative of the first invariant. So far, we failed to establish any functional relation of such a kind.

\vspace{0.5cm}

\noindent
{\bf 4. $\mathcal{N}=3$ super--Schwarzian derivative}\\

\noindent
Contracting the first invariant in (\ref{INV}) with the Levi--Civita symbol $\epsilon_{ijk}$, one reveals the $\mathcal{N}=3$ super--Schwarzian derivative introduced in \cite{Sch}
\be\label{N3S}
\mathcal{I}[\psi(t,\theta);t,\theta]:=\frac{\epsilon_{ijk} (\mathcal{D}_i \psi_l) ( \mathcal{D}_j \mathcal{D}_k \psi_l)}{\mathcal{D} \psi \mathcal{D} \psi}.
\ee
In general, an $\mathcal{N}$--extended super--Schwarzian must obey two conditions. Firstly, considering a superconformal diffeomorphism $t'=\rho(t,\theta)$, $\theta'_i=\psi_i (t,\theta)$ of ${\mathcal{S}}^{1|\mathcal{N}}$ superspace
and changing the argument of the super--Schwarzian $\psi_i (t,\theta) \to \Omega_i (t',\theta')$, $\mathcal{I}[\Omega(t',\theta');t,\theta]$ should be expressible in terms of $\mathcal{I}[\psi(t,\theta);t,\theta]$ and
$\mathcal{I}[\Omega(t',\theta');t',\theta']$. This property is known as the composition law. Secondly, setting an $\mathcal{N}$--extended super--Schwarzian to zero, one should reproduce a finite--dimensional $\mathcal{N}$--extended superconformal transformation acting in the odd sector of ${\mathcal{S}}^{1|\mathcal{N}}$ superspace.

Let us verify that both the conditions are satisfied for the expression in (\ref{N3S}). Changing the argument $\psi_i (t,\theta) \to \Omega_i (t',\theta')$ and taking into account the constraints (\ref{Const2}), which give rise to
\be
\left(\mathcal{D} \Omega \mathcal{D} \Omega \right)=\frac 13 \left(\mathcal{D} \psi \mathcal{D} \psi \right) \left(\mathcal{D}' \Omega \mathcal{D}' \Omega \right),
\ee
one gets the composition law
\be\label{TRL}
\mathcal{I}[\Omega(t',\theta');t,\theta]=\mathcal{I}[\psi(t,\theta);t,\theta]+\sqrt{\frac{\mathcal{D} \psi \mathcal{D} \psi}{3}} \mathcal{I}[\Omega(t',\theta');t',\theta'].
\ee
In particular, (\ref{N3S}) holds invariant under the transformation, provided the last term in (\ref{TRL}) vanishes. To put it in other words, setting a super--Schwarzian to zero, one determines its symmetry supergroup.

Finally, let us analyse the superfield equation $\mathcal{I}[\psi(t,\theta);t,\theta]=0$. Making use of the covariant projection method, in which components of a superfield are linked to its covariant derivatives evaluated at $\theta_i=0$, and taking into account the component decomposition (\ref{sf}), (\ref{sf1}), one gets the following restrictions on the components
\be
{\dot\alpha}_i {\dot\alpha}_j=0, \qquad u {\ddot\alpha}_i-2 {\dot u } {\dot\alpha}_i=0, \qquad u {\ddot u}-2 {\dot u}^2=0.
\ee
These determine $u(t)$ and $\alpha_i (t)$
\be
u(t)=\frac{1}{c t+d}, \qquad \alpha_i (t)=\epsilon_i+\frac{{\rm i} (\epsilon_l \kappa_l) \kappa_i}{c t +d},
\ee
where
$(c,d)$ and $(\epsilon_i,\kappa_i)$ are bosonic and fermionic parameters, respectively.
The resulting superfield $\psi_i(t,\theta)$ corresponds to a finite $OSp(3|2)$ transformation acting in the odd sector of $\mathcal{S}^{1|3}$ superspace and it correctly reduces to (\ref{tr}) in the infinitesimal limit.\footnote{In order to reproduce the infinitesimal form of the superconformal boosts entering (\ref{tr}), one sets $d=1$, considers $c$ to be small, such that $\frac{1}{1+c t}\approx  1-c t$, and identifies ${\rm i} c(\epsilon_l \kappa_k) \kappa_i$ with the infinitesimal $\kappa_i$ in (\ref{tr}). The resulting transformation is a superposition of the supersymmetry transformation, special conformal transformation parametrized by $c$ and the superconformal boost associated with ${\rm i}c (\epsilon_l \kappa_k) \kappa_i$.} A finite $OSp(3|2)$ transformation acting in the even sector of $\mathcal{S}^{1|3}$ can be found by integrating Eq. (\ref{Const2}).

\vspace{0.5cm}

\noindent
{\bf 5. Conclusion}\\

\noindent
To summarize, in this work we applied the method of nonlinear realizations to the superconformal group $OSp(3|2)$ and built invariants which involve a single fermionic superfield. In contrast to the $\mathcal{N}=0,1,2,4$ cases studied in \cite{AG1,AG,GK}, the $\mathcal{N}=3$ super--Schwarzian was shown to be a particular member of a larger set of $\mathcal{N}=3$ superconformal invariants.
The advantage of the scheme above is that it is entirely focused on the finite--dimensional supergroup $OSp(3|2)$. Nether infinite--dimensional extension, nor conformal field theory techniques, nor the analysis of central charges/cocycles were needed.

In the current literature on the Sachdev--Ye--Kitaev models with $\mathcal{N}$--extended supersymmetry, it is customary to use the $\mathcal{N}$--extended super--Schwarzian derivative when defining the effective action of the relevant $1d$ quantum mechanical system. As was demonstrated above, the $\mathcal{N}=3$ case provides more possibilities.
An interesting open problem is whether the whole set (\ref{INV}) can be used to define a tractable $\mathcal{N}=3$ supersymmetric extension of the Sachdev--Ye--Kitaev model.
Note that the expressions in (\ref{INV}) carry external vector indices with respect to $SO(3)$ realised in ${\mathcal{S}}^{1|3}$ superspace.  Extra covariant derivatives acting upon (\ref{INV}) or contracting with the Levi--Civita symbol are allowed as well. For constructing the effective actions, it suffices to consider fermionic scalar combinations (integration measure in ${\mathcal{S}}^{1|3}$ is Grassmann--odd).

\vspace{0.5cm}

\noindent{\bf Acknowledgements}\\

\noindent
The author is grateful to Sergey Krivonos for useful discussions and reading the manuscript.
This work was supported by the Russian Science Foundation grant No 19-11-00005.

\end{document}